\documentclass[conference]{IEEEtran}
\IEEEoverridecommandlockouts
\usepackage{cite}
\usepackage{amsmath,amssymb,amsfonts}
\usepackage{graphicx}
\usepackage{tikz}
\usepackage{pgfplots}
\usepackage{pgfplotstable}
\usepackage{overpic}
\usepackage{ifthen}
\usepackage{hyperref}
\pgfplotsset{compat=1.7}
\usepackage{soul}
\usepackage[font=footnotesize]{subcaption}
\usepackage{tabularx}
\usepackage[font=footnotesize]{caption}
\usepackage{textcomp}
\def\BibTeX{{\rm B\kern-.05em{\sc i\kern-.025em b}\kern-.08em
    T\kern-.1667em\lower.7ex\hbox{E}\kern-.125emX}}
\usepackage{amsmath}
\usepackage{amssymb}
\usepackage{optidef}
\usepackage{algorithm}
\usepackage{svg}
\usepackage[
  letterpaper,
  top=0.7in,
  bottom=0.7in,
  left=0.75in,
  right=0.75in,
  columnsep=0.25in
]{geometry}
\usepackage{enumitem}
\usepackage{xcolor} 
\usepackage{mathtools}
\usetikzlibrary{plotmarks}

\usepackage{algpseudocode}

\newcommand*\mb[1]{\mathbf{#1}}
\newcommand*\bs[1]{\boldsymbol{#1}}

\title{Weighted Sum Rate Maximization for ITS-Aided Arrays in Multi-User MIMO}
\author{
\IEEEauthorblockN{Robbert Beerten*, Wen Shang*, and Tugce Kobal}
\IEEEauthorblockA{Nokia Bell Labs}
\thanks{*R. Beerten was with KU Leuven, Belgium, and W. Shang was with King’s College London, UK, at the time of submission. This work was completed during their internships at Nokia Bell Labs, Cambridge. The authors thank Matthew Andrews and Ryo Koblitz for helpful discussions and technical support. This work was supported by the 6G-LEADER project, funded by the European Union, supported by Smart Networks and Services Joint Undertaking (SNS JU) (Grant 101192080).}
}

\begin{document}
\maketitle%
\begin{abstract} 
This work explores the potential of integrating an Intelligent Transmissive Surface (ITS) into an antenna array to improve beamforming performance. We show that integrating a moderate number of passive refractive elements into a small antenna array can significantly improve the Weighted Sum Rate (WSR). We investigate the optimization of the WSR under two distinct operational constraints: a Radiated Power (RP) constraint and a Transmitted Power (TP) constraint. Our analysis reveals that the choice between these constraints significantly impacts the design parameters of the ITS-aided array. By contrasting these approaches, we explore critical design and material parameters, including the array geometry, surface loss, and illumination strategies.  
\end{abstract}
\section{Introduction}

In recent years, massive multiple-input multiple-output (MIMO) systems with fully digital beamforming have attracted significant attention due to their unique properties.
The large number of antennas makes the wireless channel behave less randomly, a phenomenon known as channel hardening, and makes the wireless channels of different users virtually orthogonal, a phenomenon known as favourable propagation \cite{mimobook}.
Its deployment has rapidly evolved from a theoretical concept to practical networks, and it has become an enabling technology for 5G communications. 

However, in fully digital beamforming, each antenna requires its own digital circuitry, which we refer to as a digital front-end. These digital front-ends operating at higher frequencies consume significant power. 
To this end, the concept of hybrid front-ends has been explored. These combine digital precoding with analog beamforming. They have been moderately successful in reducing power consumption in the aforementioned higher-frequency bands. However, those front-ends require a very large number of signal dividers, combiners, and phase shifters, which significantly limit implementation cost savings and power savings. 
 
Simultaneously, Intelligent Surfaces (IS) have gained significant traction for controllable wireless environments. By serving as tunable reflecting (RIS) or refracting (ITS) scattering elements, they shape the propagation environment \cite{renzoSmartRadioEnvironments2019, wuIntelligentReflectingSurface2019}. By using a tunable, passive ITS, the system's performance can be significantly improved with minimal increase in power consumption, thereby enhancing energy efficiency \cite{huangRISEnergy2019}.
Furthermore, the low cost of the phase-shifting elements makes this approach highly attractive. 

To address these challenges, we propose embedding an ITS into a small, fully digital antenna array to boost the sum rate and energy efficiency.  
In this case, a small number of active antenna elements illuminate a large intelligent surface, which then refracts or reflects the impinging signal into (a) desirable direction(s). Recently, several works \cite{jamaliIntelligentSurfaceAidedTransmitter2021} have explored this concept for Single-User MIMO (SU-MIMO) beamforming. 
Liu \textit{et al.} \cite{liuCompactUserSpecificReconfigurable2022} have also investigated this topic by deploying multiple ITS layers at the user side, thereby improving transmitter performance in a highly cost-efficient manner. However, this approach requires digital and analog precoding not only at a single ITS but at multiple stacked ITSes.
Tunali \textit{et al.} \cite{tunaliEnergyEfficiencyMaximization2025} explored energy efficiency maximization for a transceiver with a metasurface placed in front of a small array of active antennas. Tiwari and Caire \cite{tiwariEfficientModularPragmatic2025} explore the integration of an RIS with a digital array, but do not explore the optimal RIS configuration. 
Interdonato \textit{et al.} \cite{Interdonato_Di_Murro_D’Andrea_Di_Gennaro_Buzzi_2025} investigated min-max spectral efficiency maximization for an ITS-integrated array. Jamali \textit{et al.} \cite{jamaliIntelligentSurfaceAidedTransmitter2021} consider a similar system but do not solve the multi-user WSR problem. 

\noindent \textbf{Contribution:} This work presents the following contributions: 
 
\begin{itemize}
    \item We formulate and solve the multi-user sum rate maximization problem for the ITS-aided array under two distinct regimes: Radiated Power (RP) and Transmitted Power (TP) constraints. We develop a method based on the Block Coordinate Descent (BCD) and Weighted Minimum Mean Squared Error (WMMSE) algorithms, along with a Zero-Forcing Waterfilling (ZF-WF) method, which balances computational complexity with performance.
    \item We demonstrate that the optimal geometric and hardware configuration is strongly influenced by the considered constraint. Furthermore, our results establish the operational resilience of the system, demonstrating that the ITS-aided array maintains a significant performance gain over conventional arrays even under relatively high surface losses, thereby justifying the use of cost-effective transmissive materials.
\end{itemize}

\textbf{Notation :} The complex multivariate normal distribution, with covariance $\mb{A}$ and zero mean is denoted by $\mathcal{CN}(\mb{0}, \mb{A})$.
The basis vector with index $a$ as the non-zero element is indicated by $\mb{e}_a$. 
The function $\max(0,x)$ is represented by $[x]_+$. The $i,j$'th element is selected from matrix $\mb{A}$ by $[A]_{i,j}$.

\section{System Model}
\begin{figure}
    \centering
    \includegraphics[width=0.87\linewidth]{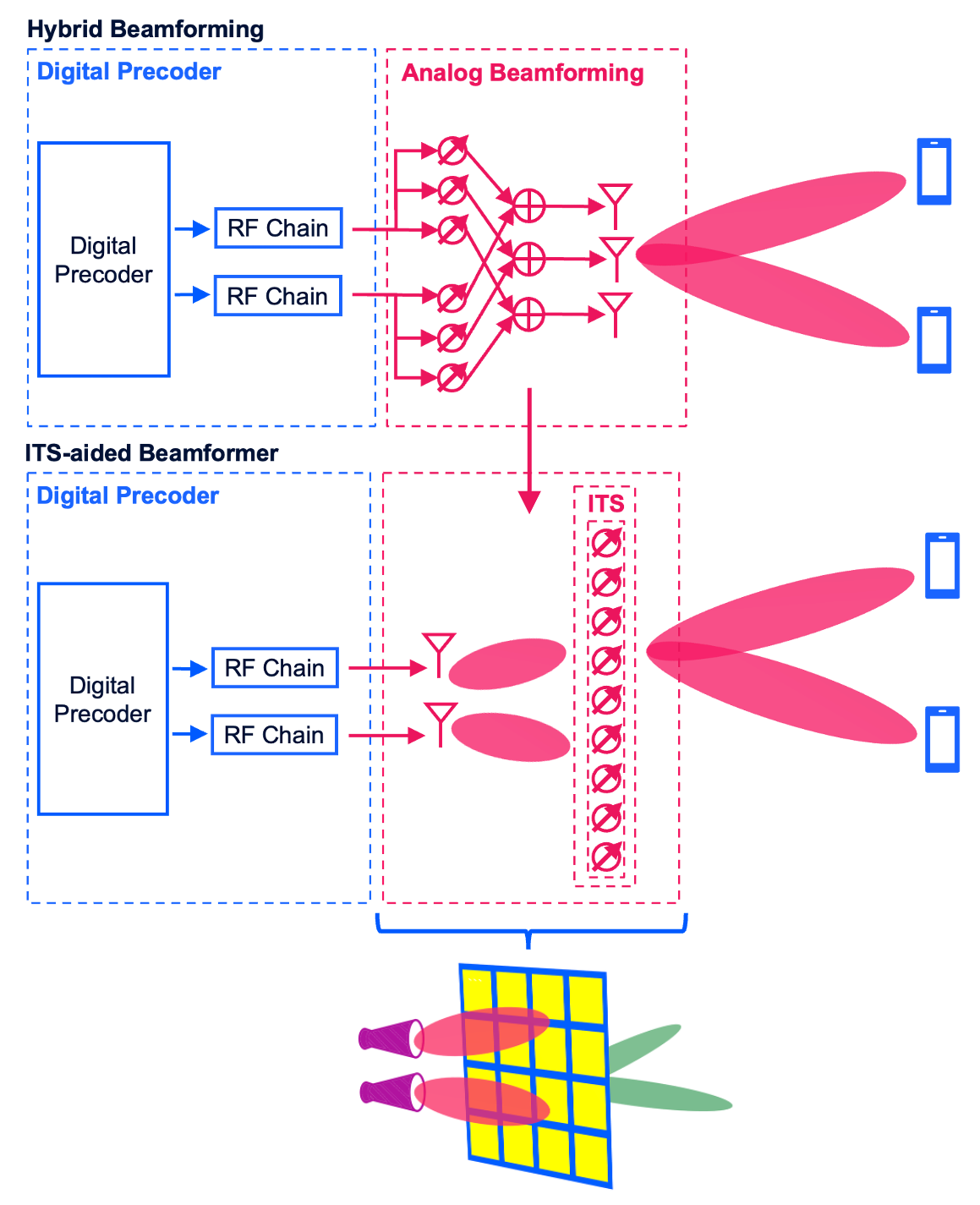}
    \caption{High-level architecture of the ITS-aided MU-MIMO beamformer, showing the system configuration with digital beamforming and ITS phase shifts. The distinction from hybrid beamforming is highlighted. \vspace{-0.5cm}}
    \label{fig:HBFtoITS}
\end{figure}

We consider a multi-user MIMO (MU-MIMO) system with a single transmitter utilizing $N$ digital RF chains and $M$ passive ITS elements, serving $K$ single-antenna users. Beamforming optimization is particularly challenging since the ITS is shared among all active antennas, and there is no direct path between the active antennas and the users.
The high-level architecture of the ITS-aided beamformer, including our system configuration and its distinction from a conventional hybrid beamforming design, is shown in Fig.~\ref{fig:HBFtoITS}.
The beamforming optimization concerns two matrices: the digital beamforming matrix $\mathbf{B}$ and the phase-shift matrix at the ITS $\mathbf{D}$. The inter-array response, $\mathbf{T}$, is assumed to be fixed upon construction of the array. Section \ref{sec:array} explains how $\mathbf{T}$ is determined. The wireless channel between the ITS-integrated array and the users is denoted by $\mb{H} \in \mathbb{C}^{K \times M}$ and is assumed to be perfectly known.
We assume that the phase shifts can be updated at the same time intervals as the digital precoding. 
The transmitted signal to all users is denoted by $\mathbf{s}$, and is assumed to be i.i.d. distributed as $\mb{s} \sim \mathcal{CN}(0, \mathbf{I}_K)$. The received signal at the $k$-th user is modeled as follows:
\begin{equation}
\begin{aligned}
    y_k &= \mathbf{h}_k^T \mathbf{D} \mathbf{T} \mathbf{B} \mathbf{s}  + n_k\\ 
    &= \mathbf{h}_k^T \mathbf{D} \mathbf{T} \sum_{k=1}^{K} \mb{b}_k s_k   + n_k \\ 
    &= \underbrace{\mathbf{h}_k^T \mathbf{D} \mathbf{T} \sum_{\substack{i = 1\\ i \neq k}}^{K} \mb{b}_i s_i}_{ \text{interference}} + 
    \underbrace{\mathbf{h}_k^T \mathbf{D} \mathbf{T}  \mb{b}_k s_k }_{\text{signal}}  + n_k,
    \end{aligned}
\end{equation}
where the thermal noise at the user is distributed as $n_k \sim \mathcal{CN}(0 ,\sigma^2)$.
Consequently, the instantaneous signal-to-interference-plus-noise ratio (SINR) for the $k$-th user is:
\begin{equation}
    \text{SINR}_k( \mb{D}, \mb{B}) =  \frac{|\mathbf{h}_k^T \mathbf{D} \mathbf{T} \mb{b}_k|^2} {\sum_{\substack{i = 1\\ i \neq k}}^{K} |  \mathbf{h}_k^T \mathbf{D} \mathbf{T}  \mb{b}_i |^2 + \sigma^2}.
 \end{equation}
The spectral efficiency (SE) of the $k$-th user is computed as:
\begin{equation}
    \text{SE}_k = \log_2(1 + \text{SINR}_k( \mb{D}, \mb{B})).
\end{equation}
This is subsequently used to define the optimization objective, the weighted sum rate (WSR) as a function of the digital beamforming matrix, $\mb{B}$, and the phase shifts of the ITS, $\mb{D}$:
\begin{equation}
    \text{WSR} = \sum_{k=1}^{K} \alpha_k \log_2(1 + \text{SINR}_k( \mb{D}, \mb{B})),
\end{equation}
where $\alpha_k$ denotes the scheduling weight of the $k$-th user.
We consider the joint optimization of $\mb{B}$ and $\mb{D}$.
This objective is maximized under two distinct power constraints, with $P_{\text{max}}$ representing the system's fixed maximum power:
\begin{itemize} 
    \item The \textbf{Radiated Power (RP)} of the ITS-aided array is limited by the constraint\footnote{We rely on a similar argument as Jamali \textit{et al. }\cite{jamaliIntelligentSurfaceAidedTransmitter2021} to argue that this constraint models the radiated power. This is true under the following assumptions: all passive elements are identical, uncoupled, and have equal transmission loss. We argue that these assumptions are reasonable and significantly improve the mathematical tractability of the constraint.}: 
    \begin{equation}
         \| \mb{D} \mb{T} \mb{B} \|_F^2  < P_{\text{max}}.
        \label{eq:eirpcons}
    \end{equation}
    \item  The \textbf{Transmitted Power (TP)} of the active antennas is limited by the constraint:
    \begin{equation}
            \| \mathbf{B}\|_F^2 < P_{\text{max}}.
            \label{eq:tpcons}
\end{equation}
\end{itemize}
While the RP constraint is relevant from a regulatory perspective, the TP constraint facilitates a more effective analysis of the array’s efficiency and the influence of its geometry. In the following discussion, the choice of constraint is kept implicit until its explicit expression is required. The function $h(\mathbf{D}, \mathbf{B})$ is introduced for abstraction:
\begin{equation}
    h(\mathbf{D}, \mathbf{B}) = \begin{cases}
        \| \mb{D} \mb{T} \mb{B} \|_F^2 & \text{RP Constraint} \\ 
         \| \mathbf{B}\|_F^2 & \text{TP Constraint}.
    \end{cases}
\end{equation}

\section{Array Model}
\label{sec:array}

The ITS passively refracts the impinging signal towards the users. 
It comprises discrete refracting elements, each independently tunable to apply a phase shift and placed apart at half wavelength, i.e. $\lambda/2$. This section describes the channel between the active antennas and the ITS. The active antennas are placed in a circle of radius $R_a$ on a plane that is symmetric with respect to the rectangular ITS. The two planes are separated by a distance $d$. Notably, Jamali \textit{et al.} \cite{jamaliIntelligentSurfaceAidedTransmitter2021} discovered that the distance $R_0 = \frac{\lambda}{2}\sqrt{\frac{M}{\pi N }}$ is significant and we use it later to dimension our experiments. 

\subsection{Active Antennas}
The array of active antennas consists of $N$ highly directive horn antennas.
Each active antenna is modeled as an axisymmetric Lambertian antenna \cite{balanis}.
Let $\theta$ denote the off-axis (boresight) angle, and $\kappa$ a parameter of the antenna gain pattern, with larger $\kappa$ indicating a more directive antenna. The gain pattern is given by \cite{balanis}:
   $ G_{\text{ant}}(\theta) = 2(1 + \kappa) \cos^{\kappa}(\theta).$

\subsection{Inter-Array Response}
Next, we model the matrix $\mathbf{T} \in \mathbb{C}^{M \times N}$, representing the channel between the active antennas and the ITS. 
The matrix also accounts for the linear transmission loss of the ITS, $\rho_{\text{ITS}}$.
This loss arises from non-idealities in the construction of the ITS. 
We assume the transmission loss is identical across the different elements. 
The channel between active element $n$ and ITS element $m$ is uniquely determined by their distance $r_{m,n}$ and the off-axis angle $\theta_{m,n}$. The path loss between active element $n$ and passive element $m$, including the loss in the ITS, is then calculated as:
\begin{equation}
    c_{m,n} = \frac{\lambda}{4 \pi r_{m,n}} \sqrt{\rho_{\text{ITS}}G_{\text{ant}}(\theta_{m,n})},
\end{equation}
and the ${m,n}$-th element of $\mathbf{T}$, representing the transmission from active antenna $n$ to passive element $m$, is:
\begin{equation}
    [\mb{T}]_{m,n} = c_{m,n} e^{-j \frac{2 \pi r_{m,n}}{\lambda}}.
\end{equation}
As noted, this matrix is fixed upon construction and depends on the relative geometry of the antennas and the ITS.


\subsection{Illumination Modes}
\label{subsec:illum}
\begin{figure}[htb]
\centering
\begin{subfigure}{.48\linewidth}
  \centering
  \includegraphics[width=.8\linewidth]{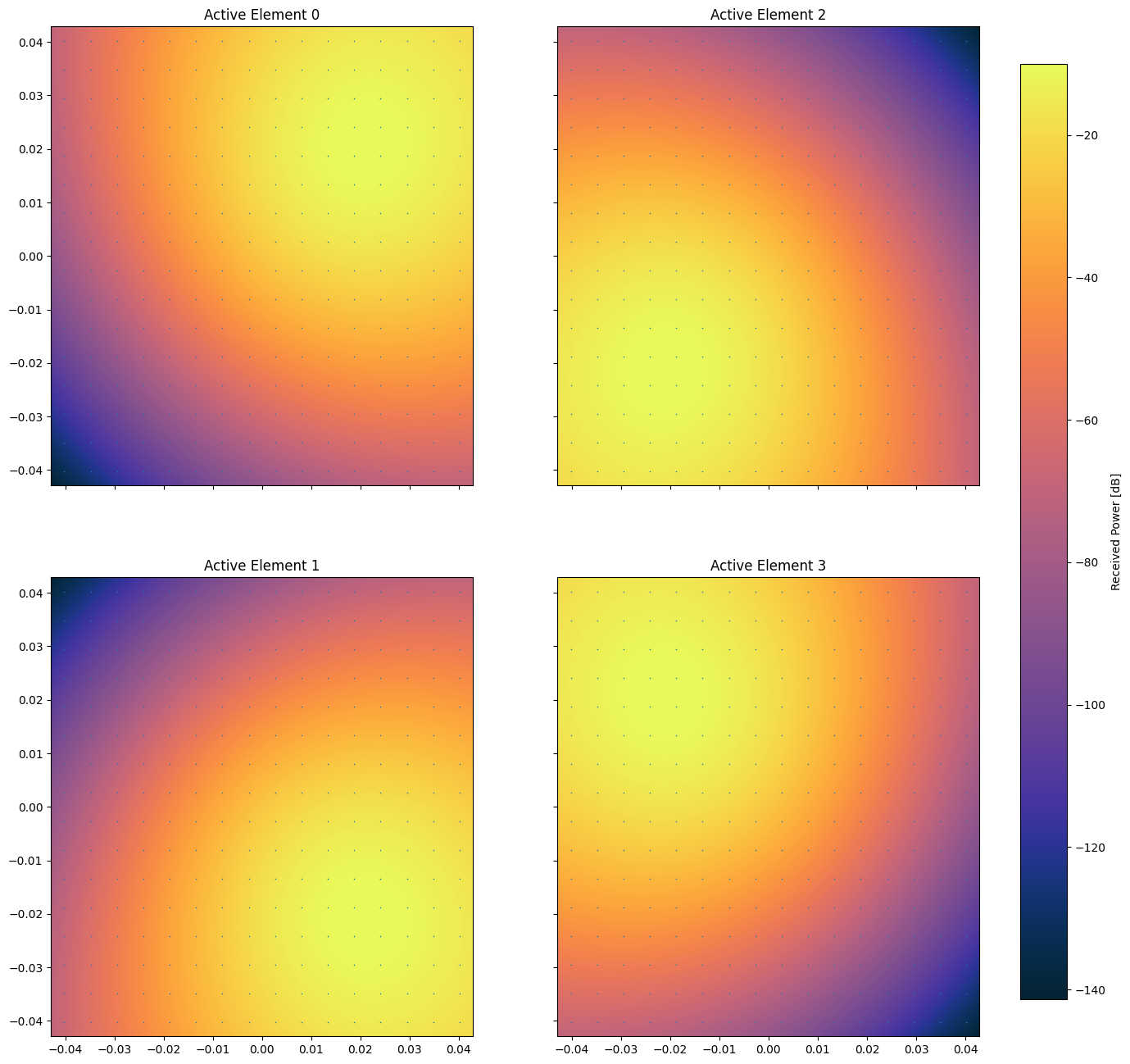}
  \caption{Partial Illumination ($\kappa$ = 49)}
  \label{fig:sub1}
\end{subfigure}%
\begin{subfigure}{.48\linewidth}
  \centering
  \includegraphics[width=.8\linewidth]{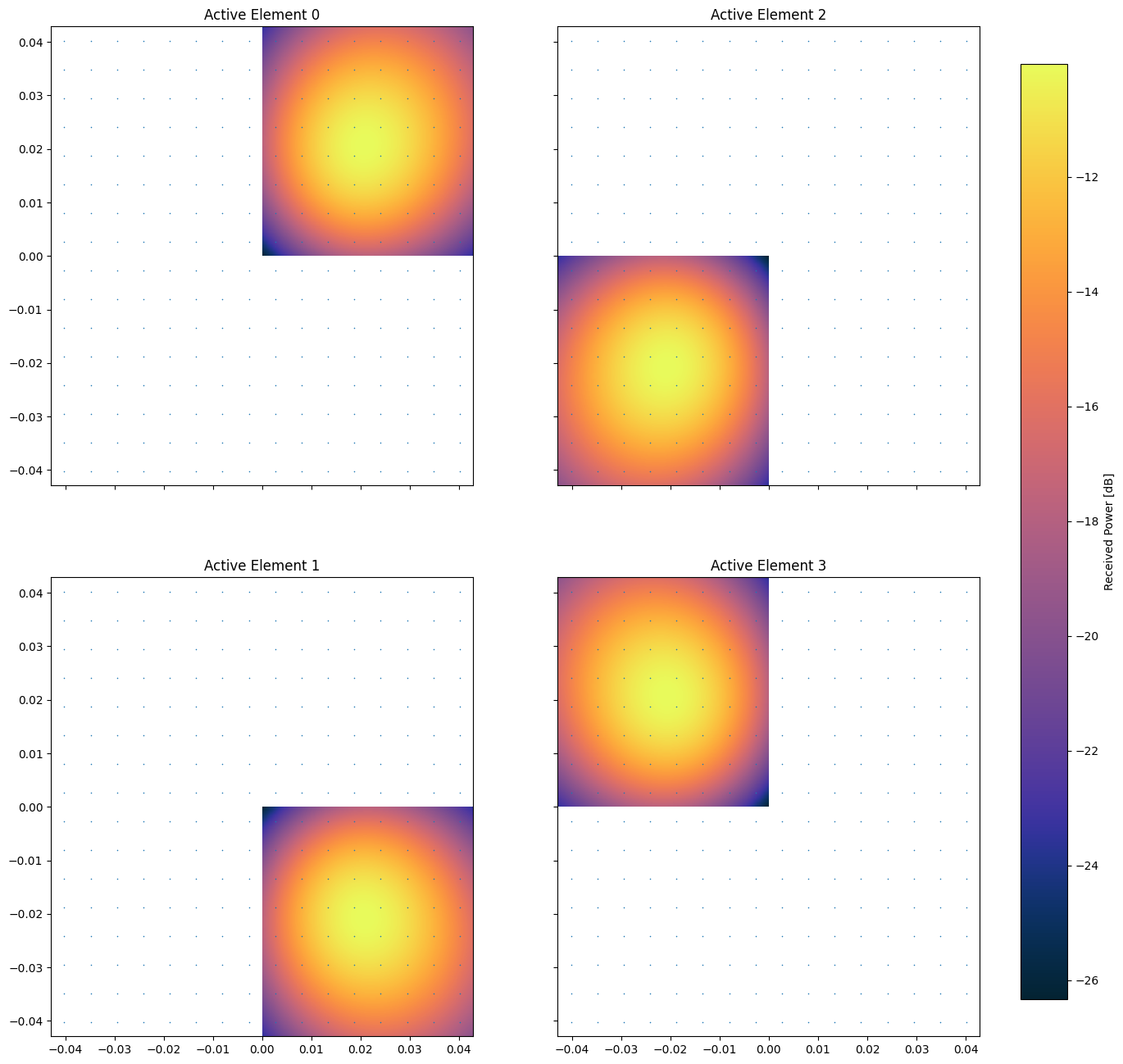}
  \caption{Separate Illumination}
  \label{fig:sub2}
\end{subfigure}
\caption{A comparison of the received power pattern at the passive surface for two illumination strategies for $R_d = 10 R_0$ and $R_a = \lambda$ from the different active antennas separately. Note that the separate illumination is unlikely to be practically achievable, but partial illumination can easily be achieved with highly directive conical horn antennas.}
\label{fig:rec_power}
\end{figure}
Based on the geometry of the active antenna array, we investigate this architecture using three illumination methods commonly discussed in the literature:
\begin{enumerate}
    \item \textbf{Full Illumination (FI):} Each active element fully illuminates the passive surface, limited only by the antenna directivity. The active antennas are oriented to cover as much of the surface as possible.
    \item \textbf{Partial Illumination (PI):} Each active element illuminates only a portion of the surface, with minimal overlap between the zones, determined by the combination of orientation and horn antenna directivity.
    \item \textbf{Separate Illumination (SI):} An idealized case in which each active element illuminates a distinct, non-overlapping portion of the passive surface.
\end{enumerate}

Fig.~\ref{fig:rec_power} provides an example of how the signals from the active antennas impinge on the passive surface under different illumination modes.
The reader is referred to \cite{jamaliIntelligentSurfaceAidedTransmitter2021} for more details on these illumination modes, which play a key role in the system’s performance.

\section{Beamforming Optimization}
This section presents our approach for jointly optimizing $\mathbf{D}$ and $\mathbf{B}$, culminating in the WSR maximization problem:
\begin{equation}
    \begin{aligned}
        \text{\textbf{P1}}: \max_{\mathbf{D},  \mathbf{B}} \quad & f_0(\mathbf{D},  \mathbf{B}) =\sum_{k=1}^{K} \alpha_k \log_2(1 + \text{SINR}_k( \mb{D}, \mb{B})) \\ 
        \text{s.t} \quad & \; h(\mathbf{D}, \mathbf{B})  \leq P_{\text{max}},
    \; | [\mathbf{D}]_{ii} |  = 1.     \end{aligned}
         \end{equation}
\vspace{-0.6cm}
\subsection{BCD-WMMSE}
We first outline a BCD method for maximizing the WSR.
Following Guo \textit{et al.} \cite{guoWeightedSumRateMaximization2020}, the joint analog and digital beamforming problem can be decomposed into two subproblems.
For the digital beamforming subproblem, we find that the solution depends on the chosen constraint. 
We first specify the exact decomposition and then introduce the subproblems for analog and digital beamforming. We introduce the following auxiliary notation for ease of exposition:
\begin{equation}
\begin{aligned}
    \text{SINR}_k( \mb{D}, \mb{B}) =  \frac{ | F_k( \mb{D}, \mb{B})|^2}{G_k(\mb{D},
    \mb{B})}.
           \end{aligned}
\end{equation}
Following \cite{shenFractionalProgrammingCommunication2018a} and \cite{guoWeightedSumRateMaximization2020}, the objective is reformulated through Lagrange and quadratic decompositions to:
\begin{equation}
\begin{aligned}
    f_1(\mb{D}, \mb{B}, \bs{\gamma}, \mb{y}) &= \sum_{k=1}^{K} \alpha_k \log_2(1 + \gamma_k) - \sum_{k=1}^{K} \alpha_k \gamma_k \\ 
        & + \sum_{k=1}^{K}  2 \sqrt{\alpha_k(1 + \gamma_k)} \text{Re} \{ y_k^{\dagger} F_k( \mb{D}, \mb{B}) \} \\ 
        & \; - \sum_{k=1}^{K} |y_k|^2 (G_k(\mb{D}, \mb{B}) + | F_k( \mb{D}, \mb{B}) |^2),
                \end{aligned}
\end{equation}
where $\gamma_k$ and $y_k$ are auxiliary variables used to separate the SINR from the logarithm and to linearize the fractional terms. The decomposed problem is then formulated with the new objective function and has the same solution as \textbf{P1} \cite{shenFractionalProgrammingCommunication2018}:
\begin{equation}
    \begin{aligned}
        \text{\textbf{P2}} : \max_{\mb{D}, \mb{B}, \bs{\gamma}, \mb{y}} \quad &  f_1(\mb{D}, \mb{B}, \bs{\gamma}, \mb{y}) \\ 
        \text{s.t} \quad & \; h(\mathbf{D}, \mathbf{B})  \leq P_{\text{max}}, \; | [\mathbf{D}]_{ii} |  = 1.
     \end{aligned}
\end{equation}
We solve this via BCD by splitting \textbf{P2} into separate problems for $ \bs{\gamma}, \mb{y}, \mb{D}$, and $ \mb{B},$. 
The subproblems for $\bs{\gamma}$ and $\mb{y}$ are solved in closed form \cite{shenFractionalProgrammingCommunication2018}: 
\begin{gather}
    \gamma_k^{\ast} = \frac{|\mathbf{h}_k^T \mathbf{D}^{\ast} \mathbf{T} \mb{b}_k^{\ast}|^2}{| \sum_{i \neq k}^{K} \mathbf{h}_k^T \mathbf{D}^{\ast} \mathbf{T}  \mb{b}_i^{\ast} |^2 + \sigma^2},  \label{eq:gamma_opt} \\ 
 y^{\ast}_k = \frac{\sqrt{\alpha_k(1 + \gamma_k) } F_k( \mb{D}, \mb{B})}{G_k(\mb{D}, \mb{B}) + | F_k( \mb{D}, \mb{B}) |^2}.
 \label{eq:y_opt}
\end{gather}
The next subsections describe our approach to solving the subproblems for $\mathbf{D}$ and $\mathbf{B}$.

\subsubsection{Analog Beamforming}
\label{subsec:analog_beamform}
We use the SCA approach proposed by Guo \textit{et al.} \cite{guoWeightedSumRateMaximization2020}.
When $\mb{B}, \bs{\gamma}$, and $\mb{y}$ are fixed, the subproblem for $\mb{D}$ can be written as\footnote{The power constraint is not addressed in this section since in Alg.~\ref{alg:cap} digital beamforming is always solved after the analog beamforming which does account for the power constraint.}:
\begin{equation}
    \begin{aligned} 
        \text{\textbf{P3}} & : \max_{\mathbf{D}} \quad  \;  
        \sum_{k=1}^{K} 2 \sqrt{\alpha_k(1 + \gamma_k)} \text{Re} \{ y_k^{\dagger} F_k( \mb{D}, \mb{B}) \} \\ 
        & - \sum_{k=1}^{K} | y_k|^2 (G_k(\mb{D}, \mb{B}) + | F_k( \mb{D}, \mb{B}) |^2) \\ 
        & \text{s.t.}  \; | [\mb{D}]_{ii} |  = 1.
     \end{aligned}
     \label{eq:beamforming}
\end{equation}
We denote the effective channel from the signal for user $i$ towards user $k$ for the ITS steering by $\mb{a}_{i,k} = \text{diag}(\mb{h}_k)  \mb{T} \mb{b}_i $. Using this definition, we introduce the following notation:
\begin{equation}
\begin{aligned}
    \bs{\nu} &= \sum_{k=1}^{K } \sqrt{\alpha_k(1 + \gamma_k)} y_k^{\dagger} \mb{a}_{k,k} \\ 
    \mb{U} &= \sum_{k=1}^{K} | y_k|^2 \sum_{i=1}^{K} \mb{a}_{k,i}\mb{a}_{k,i}^H.
    \end{aligned}
\end{equation}
The phase shift optimization is then rewritten as follows: 
\begin{equation}
    \begin{aligned} 
        \max_{\bs{\psi}} \quad & \;  f_2(\bs{\psi}) =  
         2 \text{Re} \{ \bs{\psi}^H \bs{\nu} \} -   \bs{\psi}^H \mb{U} \bs{\psi} \\ 
        \text{s.t.} & \; | \bs{\psi}_{i} |  = 1 \qquad \forall i \in \{ 0, \dots, M \},
     \end{aligned}
     \label{eq:beamforming}
\end{equation}
The problem can then be simplified by optimizing the phases, $\bs{\phi}$, directly rather than $\bs{\psi}$ \cite{guoWeightedSumRateMaximization2020}, and solved by gradient ascent. The gradient of the objective is given by: 
\begin{equation}
    \nabla_{\bs{\phi}} f_3(\bs{\phi}) =  \text{Re} \{ -j e^{-j \bs{\phi}} (\bs{\nu}  - 2 \mb{U} e^{j\bs{\phi}})\}.
\end{equation}
Furthermore, each iterate must be projected back onto the feasible region as follows: 
\begin{equation}
    \text{Proj}(\bs{\phi}) = \mb{\bs{\phi}} \mod 2 \pi
\end{equation}
The $k+1$'st gradient ascent iterate is then: 
\begin{equation}
    \bs{\phi}^{(k+1)} = \text{Proj}( \bs{\phi}^{(k)} +  \tau \nabla_{\bs{\phi}} f_3 (\bs{\phi}^{(k)})),
\end{equation}
where $\tau$ is the stepsize, which is found by performing a line search over $\tau$ until the Armijo condition is satisfied: 
\begin{equation}
\begin{aligned}
    \tau \zeta  \| \nabla_{\bs{\phi}} &f_3 (\bs{\phi}^{(k)}) \|_2^2 \leq \\
    \; &f_3\left(\text{Proj}\left( \bs{\phi}^{(k)} +  \tau \nabla_{\bs{\phi}} f_3 (\bs{\phi}^{(k)}) \right) \right) - f_3(\bs{\phi}^{(k)}), 
    \end{aligned}
\end{equation}
where $\zeta$ is a parameter that determines what constitutes a sufficient increase under the Armijo condition. 
We empirically observed that a proper choice for $\zeta$ is 0.001. 

\subsubsection{Digital Beamforming}
The digital beamforming problem can be solved similarly to the approach proposed in \cite{shiIterativelyWeightedMMSE2011a} for the RP constraint; we show that a markedly different solution strategy is required under the TP constraint.
\begin{equation}
    \begin{aligned} 
        \text{\textbf{P4}} : \max_{\mathbf{B}} \quad & \;  
        \sum_{k=1}^{K} 2 \sqrt{\alpha_k(1 + \gamma_k)} \text{Re} \{ y_k^{\dagger} F_k( \mb{D}, \mb{B}) \} \\ 
         & - \sum_{k=1}^{K} | y_k |^2 (G_k(\mb{D}, \mb{B}) + F_k( \mb{D}, \mb{B})) \\ 
        \text{s.t.} & \; h(\mathbf{D}, \mathbf{B}) \leq P_{\text{max}}.
     \end{aligned}
     \label{eq:beamforming}
\end{equation}
By solving for a stationary point for the Lagrangian of \textbf{P4}, the digital precoder for each user $k$ under the TP constraint is obtained as:  
\begin{equation}
     \mb{b}_k = \left( \sum_{i=1}^{K}  |y_i|^2  \tilde{\mb{h}}_i^{\dagger}\tilde{\mb{h}}_i^T  + \mu^{\ast} \mb{T}^H \mb{T} \right)^{-1}  \sqrt{\alpha_k(1 + \gamma_k)}  y_k \tilde{\mb{h}}_k,
\end{equation}
For the RP constraint, the solution is found as:
\begin{equation}
     \mb{b}_k = \left( \sum_{i=1}^{K}  |y_i|^2  \tilde{\mb{h}}_i^{\dagger}\tilde{\mb{h}}_i^T  + \mu^{\ast} \mb{I}_N \right)^{-1}  \sqrt{\alpha_k(1 + \gamma_k)}  y_k \tilde{\mb{h}}_k,
\end{equation}
where $\tilde{\mathbf{h}}_k$ is the effective channel for the digital beamforming for user $k$ which is found as $\tilde{\mathbf{h}}_k^T = \mb{h}_k^T \mb{D} \mb{T}$ and $\mu^{\ast}$ is the optimal dual variable. This dual variable is determined via a line search strategy that exploits the complementary slackness of the power constraint at optimality. In other words, we must compute the digital precoding as a function of the dual variable, i.e., $\mb{B}(\mu)$. First, we check if $\mb{B}(0)$ satisfies 
$h(\mathbf{D}, \mathbf{B}(0)) \leq P_{\text{max}}$, if so $\mu^{\ast} = 0$. Otherwise, we must perform a line search over $\mu$ s.t. $h(\mathbf{D}, \mathbf{B}(\mu)) = P_{\text{max}}$. This line-search is significantly simplified by the fact that $h(\mathbf{D}, \mathbf{B}(\mu))$ is monotonically decreasing in $\mu$ for both constraint cases\footnote{For the TP constraint, the proof is similar to the proof in \cite{shiIterativelyWeightedMMSE2011a}. A parallel proof can be constructed for the RP constraint, but is omitted here due to the page limitation.}, and each iteration in the line-search only requires inversion of an $N \times N$ matrix. 

Algorithm~\ref{alg:cap} summarizes the BCD procedure, and the order of the updates of $\bs{\gamma}^{(t)}, \mb{y}^{(t)},  \mb{D}^{(t)},$ and $\mb{B}^{(t)}$ at each iteration $t$. The variables $\mb{B}^{(0)}$ and $\mb{D}^{(0)}$ are initialized according to the ZF-WF approach.
\begin{algorithm}
\caption{High-Level Algorithm}\label{alg:cap}
\begin{algorithmic}
\Require  Initialize $\mb{D}^{(0)}, \mb{B}^{(0)}$  
\While{$f_0(\mb{B}^{(t)}, \mb{D}^{(t)}) - f_0(\mb{B}^{(t-1)}, \mb{D}^{(t-1)}) > \epsilon$}
\State $\bs{\gamma}^{(t)} \gets$  (\ref{eq:gamma_opt})
\State $\mb{y}^{(t)} \gets $ (\ref{eq:y_opt})
\State $\mb{D}^{(t)} \gets \arg \max \text{\textbf{P3}}$
\State $\mb{B}^{(t)} \gets \arg \max \text{\textbf{P4}}$
\EndWhile
\end{algorithmic}
\end{algorithm}%

\subsection{Low-Complexity Approach: ZF-WF}
Given the high computational cost of the BCD-WMMSE approach, we propose an alternative low-complexity ZF-WF scheme. This scheme first maximizes the received signal power in the ITS between the users and the RF chains and then applies ZF-WF for interference cancellation and power allocation. This beamforming optimization can be solved in a single iteration, unlike the WMMSE-BCD approach.

\subsubsection{Signal Power Maximization Problem}
The signal power maximization problem, formulated via effective channel gain maximization, is given as follows:
\begin{equation}
    \begin{aligned}
        \max_{\mathbf{D}} & \;  \sum_{k=1}^{K} \| \mathbf{h}_k^T \mathbf{D} \mathbf{T} \mathbf{e}_k\|_2^2 \\ 
        \text{s.t} & \; | [\mb{D}]_{ii} |  = 1,
     \end{aligned}
\end{equation}

where $\mb{e}_k$ is the standard basis vector for the $k$'th RF chain. We assume here that the choice of which RF chain belongs to which user does not significantly affect the result, since the users are all in the far-field of the array. This formulation enables us to maximize the effective channel between one RF chain and one user. The problem can be reformulated as follows: 
\begin{equation}
    \begin{aligned}
        \max_{\bs{\theta}} & \;   \| \bs{\theta}^T \sum_{k=1}^{K} \text{diag}(\mb{h}_k) \mathbf{T} \mathbf{e}_k \|_2^2 \\ 
        \text{s.t} & \; | [\bs{\theta}_{i}] |  = 1.
     \end{aligned}
\end{equation}
With a unit-norm constraint on each element, the problem reduces to a phase alignment problem, the solution of which   is given by: 
\begin{equation}
    \bs{\theta}^{\ast} =  \frac{\big(\sum_{k=1}^{K} \text{diag}(\mb{h}_k) \mathbf{T} \mathbf{e}_k\big)^{\dagger} }{\big|\sum_{k=1}^{K} \text{diag}(\mb{h}_k) \mathbf{T} \mathbf{e}_k\big|}
\end{equation}
We interpret the division as element-wise division to normalize each element individually.
\subsubsection{Zero-Forcing Water-Filling}
We compute the digital zero-forcing precoder over the effective channel, consisting of the transfer matrix $\mb{T}$, the analog precoding $\mb{D}$, and the wireless channel $\mb{H}$, denoted by:

\begin{equation}
    \tilde{\mb{H}} =  \mb{H}\mb{D}\mb{T}
\end{equation}
The zero-forcing precoder is given by the pseudo-inverse of the effective channel matrix multiplied by the power allocation matrix, as follows:
\begin{equation}
    \mathbf{B} = ( \tilde{\mb{H}}^{H} \tilde{\mb{H}})^{-1} \tilde{\mb{H}} \mb{P} = \mb{F} \mb{P},
\end{equation}
where $\mb{P}$ and $\mb{F} = [\mb{f}_1, ..., \mb{f}_K ]$ denote the power allocation matrix and the normalized precoding vector, respectively. 
Assuming perfect channel state information, the ZF precoder eliminates all interference between users, and the power allocation is found via a water-filling approach \cite{suOptimalZeroForcingHybrid2022}:
\begin{equation}
    \begin{aligned}
        \max_{\{p_k\}_{1,..,K}} \quad & \;  \sum_{k=1}^{K} \alpha_k \log_2(1 + \frac{p_k}{\sigma^2})  \\ 
        \text{s.t} \; & \; \sum_k a_k p_k = P_{\text{max}}.
     \end{aligned}
\end{equation}
Here the definition of $a_k$ depends on the type of constraint:  
\begin{equation}
    \text{RP} : a_k = \| \mb{DT}\mb{f}_k\|^2_2, \qquad \text{TP} : a_k = \| \mb{f}_k\|^2_2.
\end{equation}
The water level for this problem is given by:
\begin{equation}
    \mu =  \frac{\sum \alpha_k }{P_{\text{max}} + \sigma^2 \sum_{k=1}^{K}a_k}
    \end{equation}
The allocated power for user \(k\) is then computed as:
\begin{equation}
        p_k = \begin{bmatrix} (\frac{\alpha_k}{\mu^{\ast}} -  a_k \sigma^2) \end{bmatrix}_+, 
\end{equation}
and the power allocation matrix is constructed as $\mb{P} = \text{diag} (\sqrt{p_1}, \sqrt{p_2} ,\dots, \sqrt{p_K} )$.
This precoder is particularly appealing due to its low computational complexity, as only the water level must be determined via a 1D line search, and the precoder requires only one inversion of an $N \times N$ matrix.

\section{Numerical Results}

This section evaluates the proposed method via numerical experiments using the 28 GHz channel model by Akdeniz \textit{et al.} \cite{Akdeniz_Liu_Samimi_Sun_Rangan_Rappaport_Erkip_2014}. We illustrate the impact of the main design parameters and compare the performance of the proposed approaches. Unless otherwise stated, the simulation parameters are listed in Table \ref{table:parameters}. The results are averaged over 1000 random user locations and channels. 
\begin{table}[htb]
\centering
\begin{tabular}{|l|l|l|l|l|l|}
\hline
Parameter & Value & Parameter & Value & Parameter & Value  \\ \hline \hline
$N$ & 4 & $M$ & 128 & $K$ & 4 \\ \hline
 $\kappa$  &  49 & $d$ & 10 $R_0$ & $\rho_{\text{ITS}}$ & -3.5 dB \\ \hline
$\sigma^2$ & $10^{-7}$ &$ R_a$& $\lambda$& $f_c$ & 28 GHz\\ \hline 
\end{tabular}
\caption{Simulation Parameters}
\label{table:parameters}
\end{table}
Although we assume equal user priority ($\alpha_k=1,\quad \forall k$), the WSR framework provides a principled formulation consistent with standard user scheduling objectives. Consequently, long-term proportional fairness (PF) can be implemented as a sequence of WSR problems by assigning each user a weight equal to the inverse of their moving average rate. This formulation enables straightforward extension to more sophisticated scheduling objectives in future work.
To demonstrate the performance of the proposed methods, we evaluate the ZF-WF and WMMSE-BCD precoding strategies across three illumination strategies: FI, PI, and SI. Performance is compared against two baselines: digital beamforming WMMSE without the ITS (No ITS) and WMMSE with random phase shifts under FI (Random).




\begin{figure}[htb]
\begin{overpic}{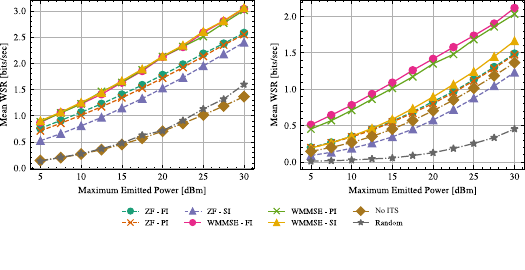}
\put(10,3){\footnotesize (a) RP Constraint}
  \put(65,3){\footnotesize (b) TP Constraint}
  \end{overpic}\vspace{-0.2cm}
  \caption{Mean WSR vs. the power constraint $P_{\text{max}}$ for $M=128$ and $N=4$ for the two different power constraints.}
  \label{fig:power}
      \vspace{-0.2cm}
\end{figure}
Fig.~\ref{fig:power} illustrates the significance of the constraint. Because there is no direct link between the active antennas and the users, efficient use of the ITS is essential.
Random ITS configuration offers no gain under the RP constraint and can degrade performance below the level achievable without an ITS under the TP constraint due to the inherent loss in the ITS, thus highlighting the need for careful joint optimization.
Strikingly, under the RP constraint, the ZF-WF method performs nearly as well as the WMMSE-BCD method. However, under the TP constraint, the ZF-WF method performs notably worse. We suspect this is due to the ZF-WF method using all the power budget and degrees of freedom to cancel all interference, whereas WMMSE-BCD can use the power budget more efficiently. Furthermore, the RP constraint achieves a higher mean WSR than the TP constraint for the same numerical values. This is because performance under the RP constraint does not suffer from surface losses in the ITS. 
We conclude that the TP constraint significantly benefits from the more precise joint optimization in the BCD-WMMSE.

\begin{figure}[htb]
\begin{overpic}{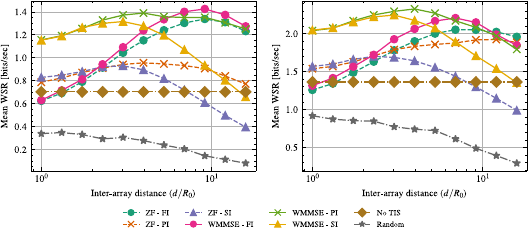}
\put(10,-1.1){\footnotesize (a) $P_{\text{max}}$ = 20dBm}
  \put(65,-1.1){\footnotesize (b) $P_{\text{max}}$ = 30dBm}
  \end{overpic}\vspace{0.1cm}
  \caption{Mean WSR vs. the inter-array distance for $N=4$ 
    and $M=128$ and a fixed surface loss of 3.5 dB for two different TP constraints.}
    \label{fig:array_dist}
    \vspace{-0.2cm}
\end{figure}

Fig~\ref{fig:array_dist} shows the effect of the inter-array distance relative to $R_0$. The optimal inter-array distance varies significantly across illumination methods. 
The ZF-WF consistently trails the WMMSE-BCD,
with the performance gap widening as the inter-array distance increases.
Additionally, the performance gap decreases with higher power constraints. Furthermore, the performance gap is smaller for the FI than for the other illumination methods. Intuitively, this means that both methods perform well when the total incident power on the ITS is high.
This can be attributed to the observation that, in fully digital beamforming, ZF-WF achieves WSRs comparable to those of WMMSE only at high SNR, and the performance gap generally increases at lower SNR. 



\begin{figure}[htb]
\begin{overpic}{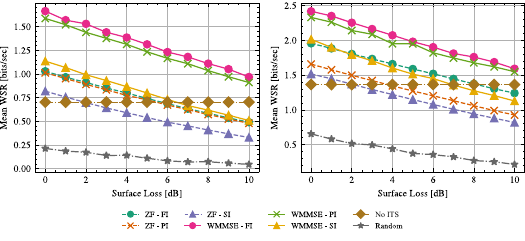}
\put(10,-1.1){\footnotesize (a) $P_{\text{max}}$ = 20dBm}
  \put(65,-1.1){\footnotesize (b) $P_{\text{max}}$ = 30dBm}
  \end{overpic}\vspace{0.1cm}
  \caption{Mean WSR vs. the linear surface loss $\rho_\text{ITS}$ for $M=128$ and $N=4$ and an inter-array distance of  $10R_0$ for two different TP constraints.}
  \label{fig:surf_loss}
  \vspace{-0.1cm}
\end{figure}

Fig.~\ref{fig:surf_loss} illustrates the impact of the surface loss, $\rho_{\text{ITS}}$.
This result is promising for the ITS-integrated array, indicating that the WSR can be significantly improved compared to an array without ITS.
The WMMSE-BCD method outperforms the system without ITS for a surface loss higher than 10 dB under FI and PI. Even for the ZF-WF solution, the ITS-integrated array outperforms the array without ITS by up to 5 dB in surface loss under FI and PI. 
Interestingly, under the TP constraint, 
a randomly configured ITS is significantly outperformed by an array without ITS in all cases.
Notably, this occurs even if the ITS is assumed to be lossless. This phenomenon can be explained by the absence of a direct channel between the user and the active antennas.

\section{Conclusion} \label{sec:conclusion} In this paper, we investigated the integration of an ITS into a small active antenna array for a multi-user MIMO system. We outlined an optimization framework for WSR maximization under two distinct power constraints: RP and TP. We have presented two optimization methods, WMMSE-BCD and ZF-WF.  Our results highlight the importance of the power constraint, the array's physical geometry, and hardware non-idealities. Notably, we demonstrated that the proposed architecture is robust to hardware non-idealities; even under high surface-loss conditions, the ITS-aided array provides substantial WSR improvements over the baseline.

\bibliographystyle{ieeetr} 
\bibliography{references} 
\end{document}